\documentclass[twocolumn]{webofc}

\usepackage[varg]{txfonts}   
\usepackage{hyperref}
\usepackage{url}
\usepackage{fancyhdr}
\usepackage{comment}
\usepackage[dvipsnames]{xcolor}

\hypersetup{colorlinks=true,citecolor=blue,urlcolor=blue,linkcolor=blue}

\newcommand{\version}{{\sc 2.0}}
\newcommand{\freya}{{\sc freya}}
\newcommand{\cgmf}{{\sc cgmf}}
\newcommand{\fifrelin}{{\sc fifrelin}}
\newcommand{\yahfc}{{\sc yahfc}}
\newcommand{\talys}{{\sc talys}}
\newcommand{\feta}{{\sc feta}}

\newcommand{\code}[1]{{\color{Blue}\texttt{#1}}}

\newcommand{\Zf}{Z_{\rm f}}
\newcommand{\Nf}{N_{\rm f}}
\newcommand{\Af}{A_{\rm f}}

\newcommand{\probIni}{\mathbb{P}_{0}}
\newcommand{\probInd}{\mathbb{P}_{\rm ind.}}

\newcommand{\Yind}{Y_{\rm ind.}}

\newcommand{\TKE}{{\mathrm{TKE}}}

\begin{document}

\thispagestyle{fancy}

\title{Fission Evaluation Tools and Analytics (FETA)}

\author{\firstname{Nicolas} \lastname{Schunck}\inst{1}\fnsep\thanks{\email{schunck1@llnl.gov}}}

\institute{Nuclear data and Theory Group, Nuclear and Chemical Science Division, Lawrence Livermore National Laboratory, California, USA 94550}

\abstract{We present a new Python package called {\feta} to compute fission
observables. In comparison to existing fission event modeling codes on the
market, {\feta} is a modular, flexible, and code-agnostic wrapper around a
statistical reaction theory program. In this paper, we report on version
{\version} of \feta, which simulates the prompt decay of the fragments with the
LLNL-developed Hauser-Feshbach code \yahfc. We discuss the design philosophy of
\feta, present some validation results for the neutron-induced fission of
$^{239}$Pu, and briefly outline the structure of the package.}

\maketitle


\section{Introduction}
\label{sec:intro}

It is difficult not to overstate the importance of nuclear fission in modern
science and technology. On a theoretical level, a predictive description of
why and how an atomic nucleus splits into two or more fragments remains a
scientific and computational grand challenge \cite{schunck2022theory}. On a
practical level, fission is what terminates nucleosynthesis of heavy elements
\cite{cowan2021origin} and drive the stability of superheavy elements
\cite{giuliani2019colloquium}, but it is also what power nuclear reactors.

\begin{figure}[!htb]
\centering
\includegraphics[width=\linewidth,clip]{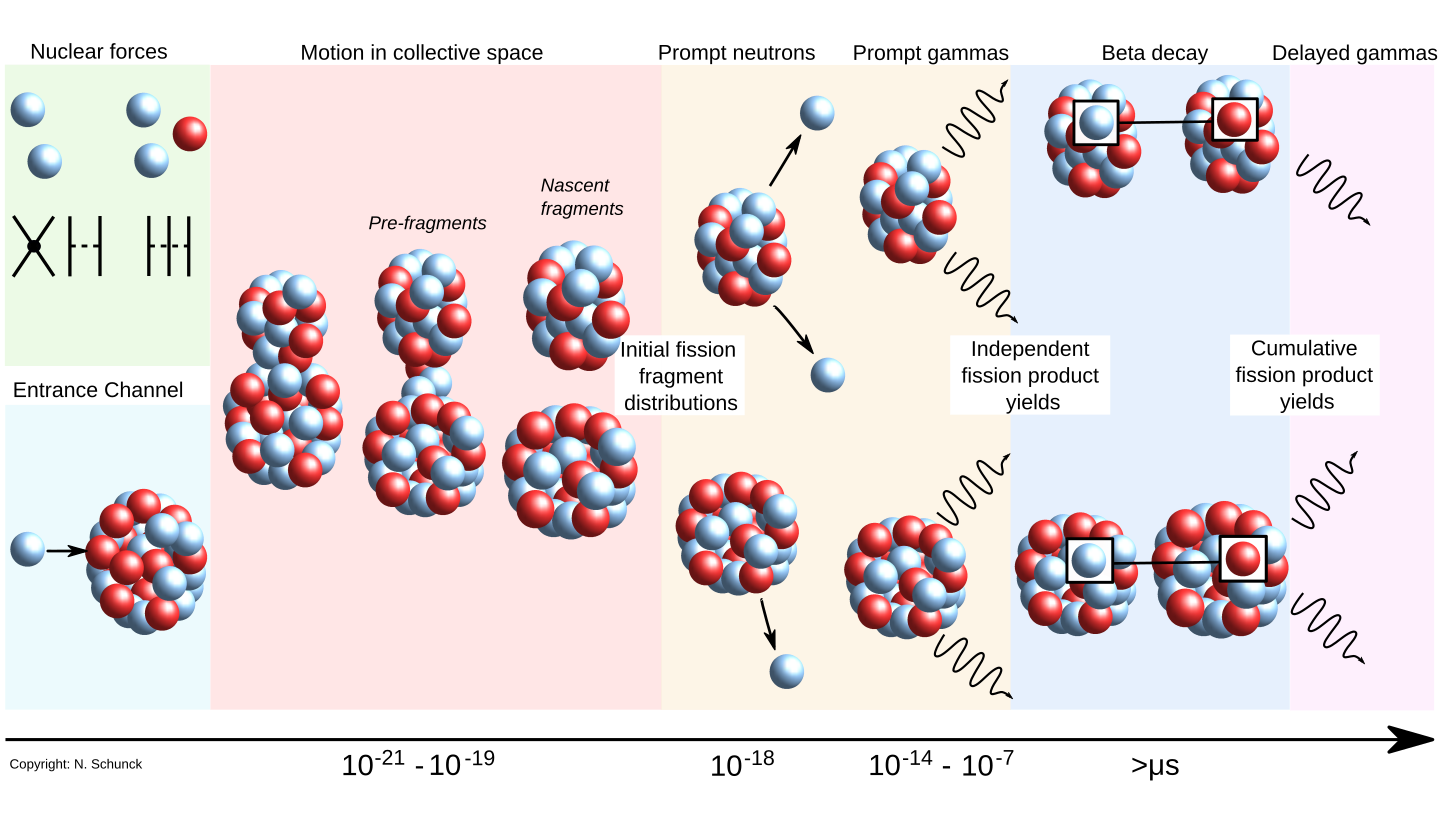}
\caption{Schematic of the fission process: the nucleus deforms itself up to the
point of scission, leading to the formation of highly excited fission
fragments that deexcite by neutron and $\gamma$ emission before $\beta$
decay and delayed emission. Each phase of deexcitation modifies the
distribution of the fission products.}
\label{fig:fission}
\end{figure}

Fission is a multistage process, as shown in Fig.\ref{fig:fission}. Following
the insights of Bohr and Wheeler \cite{bohr1939mechanism}, it is described as
an extreme deformation process that ultimately results in the formation of two
highly excited fission fragments. These fragments emit first neutrons and
photons (prompt emission) before undergoing a sequence of $\beta$ decays
followed by additional emission of (mostly) photons (delayed emission). As
the approximate timescales shown in Fig.\ref{fig:fission} suggest,
experimental measurements will become increasingly difficult as one approaches
back in time to the actual scission point. Yet, all the fission products --
fragments, neutrons, photons, electrons, antineutrino, etc. -- depend
critically on the characteristics of this scission point, which is entirely
provided by theoretical models.

Until about 20 years ago, explicitly modeling the decay of hundreds of fission
fragments was computationally too demanding. As a result, it was extremely
difficult to confront fundamental nuclear theory with actual fission
observables. For example, fission barriers, which are not observables but
extracted from fission cross section modeling, were often used as metric to
asses the quality of a theoretical model. The advent of Monte-Carlo codes
capable of following the decay of all fragments opened up new possibilities
\cite{lemaire2005monte}. Because phenomenological structure models are
inexpensive and contain adjustable parameters, they became the default choice
for most of the Monte Carlo codes on the market \cite{litaize2015fission,
verbeke2018fission,talou2021fission}. Yet, the renaissance of nuclear fission
theory in the last 10 years \cite{schunck2016microscopic} and the overall
progress in nuclear structure and reaction theory, is a strong incentive to
revisit these models. This observation is what motivated the development of the
Fission Evaluation Tools and Analytics (\feta) framework: it was designed with
the explicit goal that every single input to a set of fragment deexcitation
calculations should potentially be provided independently by the user, thereby
allowing for a systematic quantification of uncertainties from structure,
reaction and fission models.

This article is organized as follows. Section \ref{sec:design} outlines the
basic design choices of the framework. Section \ref{sec:validation} illustrates
the current capabilities on a set of standard fission observables for
$^{239}$Pu(n,f). Technical details about the framework are given in Section
\ref{sec:package} before a short conclusion.


\section{Design}
\label{sec:design}

The mission statement for {\feta} is to provide a flexible, modular,
user-friendly and efficient computational framework to compute fission
observables. Such observables include fission yields (independent, cumulative,
chain), the prompt neutron and photon spectrum, and the delayed spectrum. The
two main use cases of {\feta} are nuclear data evaluations and tests of
theoretical models of structure, reactions, decays or fission.

The main existing fission modeling codes are {\cgmf} \cite{talou2021fission},
{\freya} \cite{verbeke2018fission} and {\fifrelin} \cite{litaize2015fission}.
In contrast to them, {\feta} does not implement itself a statistical reaction
theory model but outsources the deexcitation of fission fragments to an
external program. By construction, {\feta} is thus nothing but a wrapper that
organizes the decay of a list of pairs of excited nuclei and extracts from the
outputs the global fission observables that characterize the process.
Currently, the only external program explicitly supported in {\feta} is
{\yahfc} \cite{ormand2021montea}. Adding another code would simple require
writing a pair of Python classes -- one for the input that this code needs, the
other to parse the outputs. The minimal requirement for such reaction code is
that it can decay an initial nucleus with excitation energy $E^*$, spin $J$,
and parity $\pi$ by emitting all the relevant particles, until no more decay is
possible. Codes such as {\talys} \cite{koning2023talys} could thus be
integrated in {\feta} in the future.


The initial values of $(E^{*}, J^{\pi})$ and probability $\probIni(\Zf,\Nf|Z,N)$
for every fission fragment $(\Zf,\Nf)$ given a fissioning nucleus $(Z,N)$ must
be provided by theoretical {\em fission models}. If $p_{zn}(Z,N)$ is the
probability that the nucleus $(Z,N)$ decays through a channel $c$ that
decreases the number of protons by $z$ and the number of neutrons by $n$, then,
the probability of observing the fragment $(\Zf,\Nf)$ after emission of
particles from the parent nucleus is
\begin{multline}
\probInd(\Zf,\Nf|Z,N) = \probIni(\Zf,\Nf|Z,N)  \\
  +  \sum_{z=0}^{N_z}\sum_{n=0}^{N_n} \probIni(\Zf+z,\Nf+n) p_{zn}(\Zf+z,\Nf+n) \\
  -  \sum_{z=0}^{N_z}\sum_{n=0}^{N_n} \probIni(\Zf,\Nf) p_{zn}(\Zf-z,\Nf-n) \, ,
\label{eq:Pind}
\end{multline}
where the maximum number of protons and neutrons emitted must be such that
$$
\begin{array}{rlcrl}
\max\Zf + N_z & \leq Z & & \max\Nf + N_n & \leq N \\
\min\Zf - N_z & \geq 1 & & \min\Nf - N_n & \geq 1
\end{array}
$$
The channel $n=z=0$ corresponds to the $\gamma$ channel of the fragment
$(\Zf,\Nf)$ itself. {\feta} requires that particle emission probabilities
$p_{zn}(Z,N)$ are computed and tabulated by the external reaction theory code
(here: {\yahfc}) since it will use Eq.(\ref{eq:Pind}) to combine them with the
user-provided initial fission fragment distribution $\probIni(\Zf,\Nf|Z,N)$ to
extract independent yields $\Yind(\Zf,\Nf) = 100\times \big[
\probInd(\Zf,\Nf|Z,N) + \probInd(Z-\Zf,N-\Nf|Z,N) \big] $.

The probabilities $p_{zn}(Z,N)$ themselves are computed (by the external
program) by treating each fission fragment as an excited nucleus characterized
by some energy spectrum and decay modes. The goal of {\em structure and
reaction models} is to provide quantities such as: the ground-state binding
energy; the discrete energy levels (low-energy spectrum) and the level density
(high-energy spectrum); the intensity of discrete electromagnetic transitions
(low-energy spectrum) and the $\gamma$-strength functions (high-energy
spectrum); the transmission coefficients of any light particles such as $p$,
$d$, $t$, $\alpha$, etc. This information must be supplied explicitly, either
in the form of the names of preset models, or in terms of actual data files.
It will be passed on to the external program (here: {\yahfc}) that {\feta} runs
in the background.


\section{Validation Results}
\label{sec:validation}

This section shows validation results of {\feta} for the canonical case of the
neutron-induced fission of $^{239}$Pu. We consider two different incident
neutron energies, $E_n = 0.0253$ eV (thermal), and $E_n = 14$ MeV. As mentioned
in Sec.\ref{sec:design}, the inputs from fission models include the pre-neutron
emission distributions $Y(Z,A)$, and the excitation energy and spin
distribution of all fission fragments. The mass distributions $Y(A)$ are taken
from the {\cgmf} five-Gaussian fit; the charge probabilities $p(Z|A)$ are given
by the Wahl systematics \cite{wahl1988nuclearcharge}. The excitation energy
$E^{*}$ of fission fragments is obtained by sharing the total excitation
energy, itself obtained from the conservation of energy and the value of the
total kinetic energy $\TKE$, according to the {\cgmf} prescription. All
structure or reaction model inputs are set to the ``default'' {\cgmf} option
available from \cite{talou2021fission}.

\begin{figure}[!h]
\centering
\includegraphics[width=0.959\linewidth,clip]{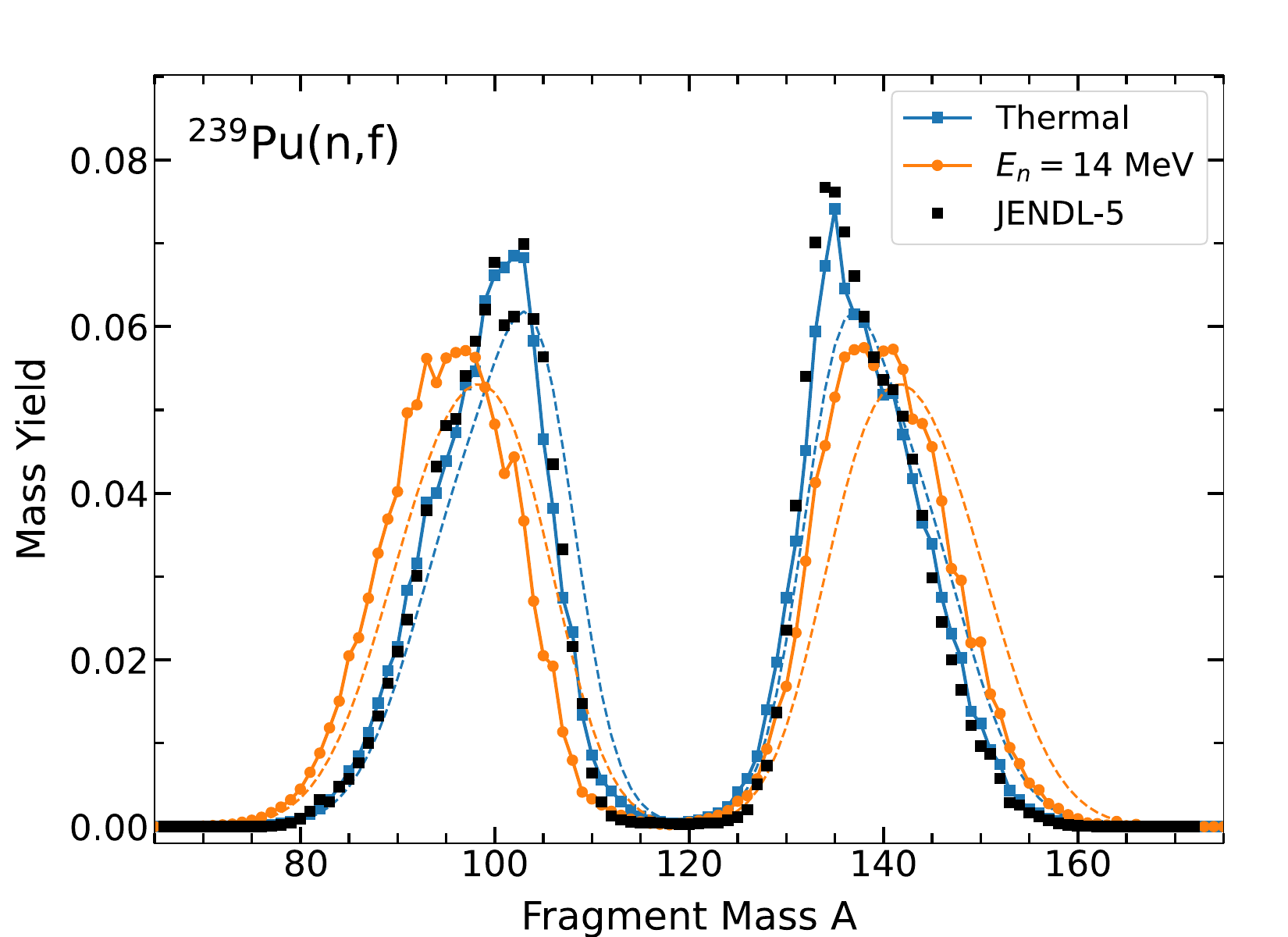}
\caption{Independent yields $\Yind(A)$ as a function of fragment mass $A$ for
$^{239}$Pu(n,f): {\feta} results for $E_n = 0.0253$ eV  and $E_n = 14$ MeV are
shown by the plain blue line with squares and plain orange line with circles,
respectively. They are compared with
the JENDL evaluation \cite{iwamoto2023japanese} (black squares). The initial
fragment mass distribution is shown by the dashed curve with no symbol.}
\label{fig:YA}
\end{figure}

As a sanity check, we show in Figure \ref{fig:YA} the independent mass yields
$\Yind(\Af) = \sum_{n=0}^{N} \Yind(\Zf=\Af-n,\Nf=n)$ as a function of the
fragment mass for two two neutron energies considered. They are compared with
the JENDL evaluation \cite{iwamoto2023japanese} and show very reasonable
agreement.

\begin{figure}[!h]
\centering
\includegraphics[width=0.99\linewidth,clip]{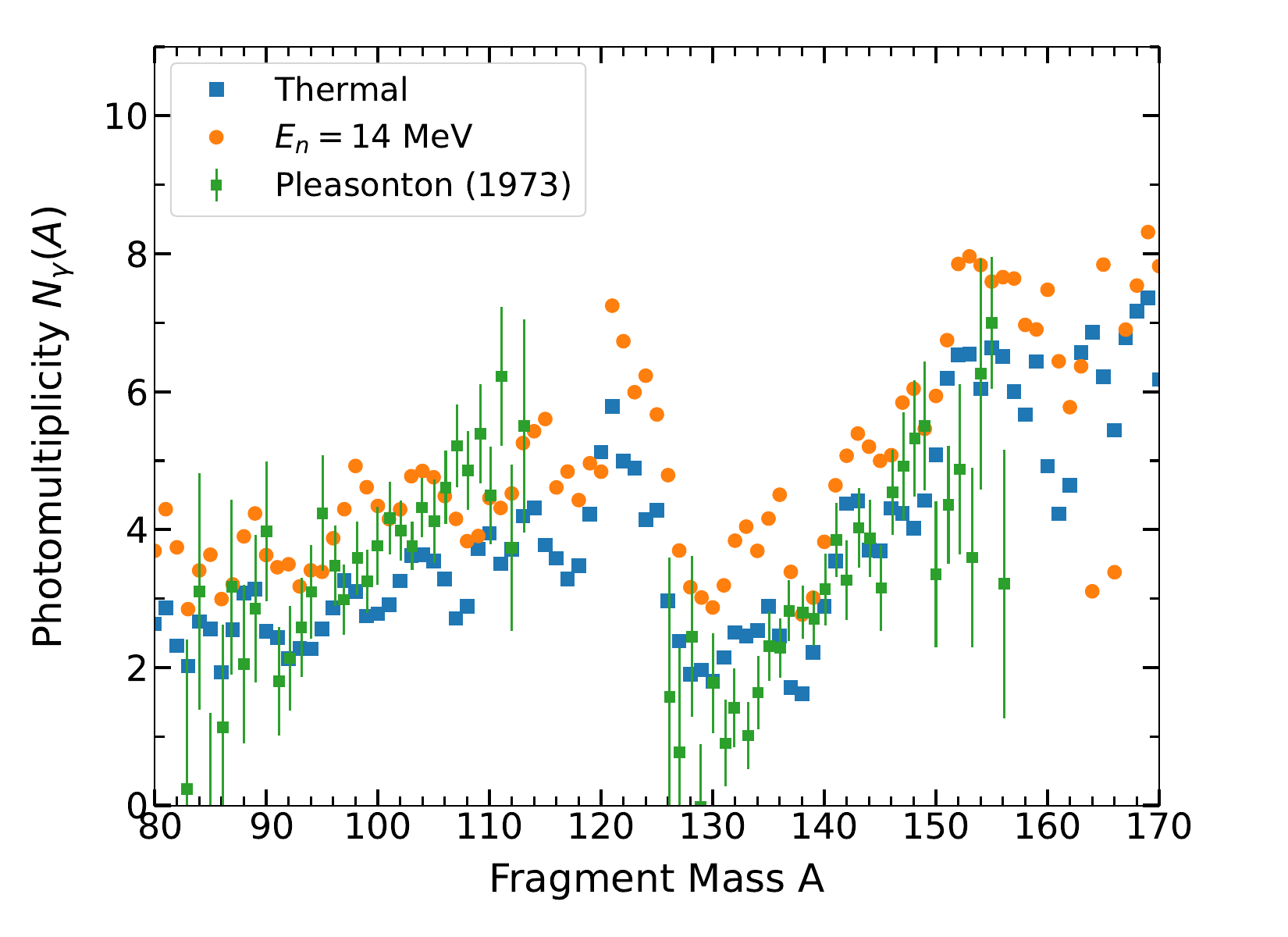}
\caption{Average number of photons per fragment mass $N_{\gamma}(A)$ for
$^{239}$Pu(n,f): {\feta} results for $E_n = 0.0253$ eV  and $E_n = 14$ MeV are
shown by the plain blue line with squares and plain orange line with circles,
respectively. They are compared with experimental data (symbols with error
bars) from \cite{pleasonton1973prompt}.
}
\label{fig:Ngamma}
\end{figure}

To illustrate the versatility of {\feta}, we show in Fig.\ref{fig:Ngamma} the
average number of emitted photons per fragment mass. In this specific
calculation, the spin distribution was input directly in the form of a CSV file
containing the results of joint angular-momentum and particle-number projection
folded with the time-dependent generator coordinate method published in
\cite{marevic2026microscopic}. In this case, calculations with phenomenological
spin distributions and microscopic ones show remarkable agreement; see also
P. Marevic {\it et al.} in this volume.

\section{The FETA Package}
\label{sec:package}

With the exception of fission cross section, fission observables can only be
extracted after simulating the decay of all fission fragments, be it the prompt
decay (emission of prompt neutrons or photons) or the $\beta$ decay.
Specifically:
\begin{itemize}
\item Fission fragments are nothing but a collection of excited atomic nuclei
characterized by specific observables (energy, spin, parity) and some initial
probability.
\item The determination of independent yields through Eq.(\ref{eq:Pind}) simply
means transforming each fragment probability from an initial to a final value.
\item Simulating the decay of an excited nucleus requires both a set of initial
conditions but also a set of nuclear models to characterize the structure of
the nucleus: level density, $\gamma$s-strength functions and particle-emission
transmission coefficients.
\item Computing the decay of hundreds of fission fragments is a good example of
a naturally parallel problem to be handled by MPI.
\end{itemize}
Object-oriented programming with class inheritance is abundantly used in
{\feta} to implement these concepts. Python abstract base classes describe mass
tables, yields, energies, and decays, with specialized subclasses such as, e.g.,
\code{IndependentYields()} or \code{PromptDecay()} to model more specific
objects. A class \code{TaskManager()} handles all {\yahfc} calculations in
parallel through the \code{mpi4py} package.

\begin{figure}[!h]
\includegraphics[width=0.99\linewidth]{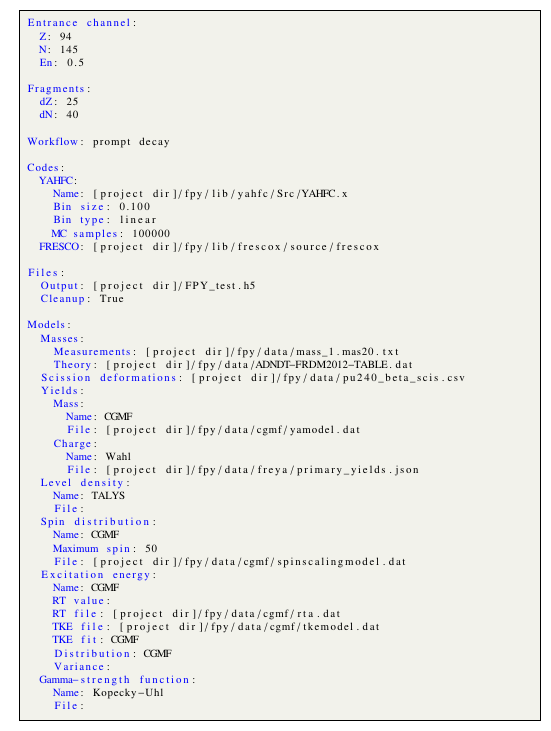}
\caption{Example of a {\feta} YAML configuration file.}
\label{fig:yaml}
\end{figure}

All {\feta} options are stored in a single YAML file. Most of these options
are related to the nuclear structure, reaction and fission input data needed by
{\feta} to deexcite fission fragments and point to files containing the data.
An example of the configuration file is provided in Fig.\ref{fig:yaml}.

\begin{figure}[!h]
\includegraphics[width=0.95\linewidth]{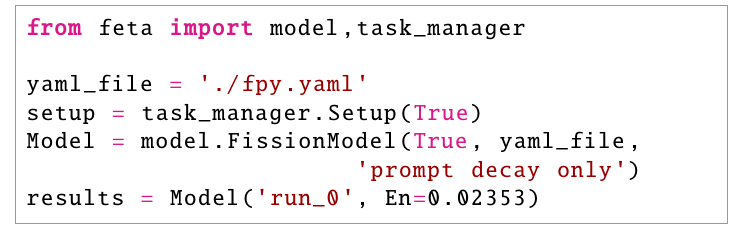}
\caption{Minimal working example of a Python script running {\feta} to compute
the independent yields and the prompt fission spectrum. In this case,
calculations are run in the directory \code{run\_0/} with MPI parallelism
enabled.}
\label{fig:run}
\end{figure}

The main user-facing {\feta} class is \code{FissionModel()}. Instantiating this
class sets up the MPI environment, defines some directories and reads the
default YAML configuration file. An instance of \code{FissionModel()} is
callable and can be passed arguments. A minimal example of running {\feta} is
the small Python script shown in Fig.\ref{fig:run}. MPI parallelism is
completely transparent to end-users: this example script could either be run
serially or in parallel through a command such as \code{mpirun} and any given
number of MPI ranks.

All results of {\feta} are stored in an HDF5 file for maximum portability,
readability and programming efficiency. A single HDF5 contains a copy of the
YAML file, the charge, mass and isotopic yields before and after prompt
emission, and the probabilities, multiplicities and energy spectra of all
emitted particles for each single fragment.


\section{Conclusion}

This article gives a very brief overview of the Python package {\feta} to
simulate the deexcitation of fission fragments and extract fission observables
from it. We emphasize that {\feta} is as flexible as the underlying reaction
theory code it uses as a computation engine: if the latter does not provide
enough flexibility to change the structure or reaction models (because such
choices are hard-coded for example), then {\feta} will be subject to the same
constraints.


\vspace{0.2cm}

\begin{acknowledgement}
This work was partly performed under the auspices of the US Department of
Energy by the Lawrence Livermore National Laboratory under Contract No.
DE-AC52-07NA27344. Computing support came from the Lawrence Livermore National
Laboratory Institutional Computing Grand Challenge program. Support for this
work was partly provided through Scientific Discovery through Advanced
Computing (SciDAC) program funded by U.S. Department of Energy, Office of
Science, Advanced Scientific Computing Research and Nuclear Physics.\smallskip

\noindent LLNL Release Number: LLNL-PROC-2019256
\end{acknowledgement}


\bibliography{zotero_output}

@article{bohr1939mechanism,
  title = {The Mechanism of Nuclear Fission},
  author = {Bohr, Niels and Wheeler, John Archibald},
  year = 1939,
  journal = {Phys. Rev.},
  volume = {56},
  number = {5},
  pages = {426},
  doi = {10.1103/PhysRev.56.426}
}

@article{cowan2021origin,
  title = {Origin of the Heaviest Elements: {{The}} Rapid Neutron-Capture Process},
  author = {Cowan, John J. and Sneden, Christopher and Lawler, James E. and Aprahamian, Ani and Wiescher, Michael and Langanke, Karlheinz and {Mart{\'i}nez-Pinedo}, Gabriel and Thielemann, Friedrich-Karl},
  year = 2021,
  journal = {Rev. Mod. Phys.},
  volume = {93},
  number = {1},
  pages = {015002},
  publisher = {American Physical Society},
  doi = {10.1103/RevModPhys.93.015002}
}

@article{giuliani2019colloquium,
  title = {Colloquium: {{Superheavy}} Elements: {{Oganesson}} and Beyond},
  author = {Giuliani, S. A. and Matheson, Z. and Nazarewicz, W. and Olsen, E. and Reinhard, P.-G. and Sadhukhan, J. and Schuetrumpf, B. and Schunck, N. and Schwerdtfeger, P.},
  year = 2019,
  journal = {Rev. Mod. Phys.},
  volume = {91},
  number = {1},
  pages = {011001},
  doi = {10.1103/RevModPhys.91.011001}
}

@article{iwamoto2023japanese,
  title = {Japanese Evaluated Nuclear Data Library Version 5: {{JENDL-5}}},
  author = {Iwamoto, Osamu and Iwamoto, Nobuyuki and Kunieda, Satoshi and Minato, Futoshi and Nakayama, Shinsuke and Abe, Yutaka and Tsubakihara, Kohsuke and Okumura, Shin and Ishizuka, Chikako and Yoshida, Tadashi and Chiba, Satoshi and Otuka, Naohiko and Sublet, Jean-Christophe and Iwamoto, Hiroki and Yamamoto, Kazuyoshi and Nagaya, Yasunobu and Tada, Kenichi and Konno, Chikara and Matsuda, Norihiro and Yokoyama, Kenji and Taninaka, Hiroshi and Oizumi, Akito and Fukushima, Masahiro and Okita, Shoichiro and Chiba, Go and Sato, Satoshi and Ohta, Masayuki and Kwon, Saerom},
  year = 2023,
  journal = {J. Nucl. Sci. Technol.},
  volume = {60},
  number = {1},
  pages = {1},
  publisher = {Taylor \& Francis},
  doi = {10.1080/00223131.2022.2141903}
}

@article{koning2023talys,
  title = {{{TALYS}}: Modeling of Nuclear Reactions},
  author = {Koning, Arjan and Hilaire, Stephane and Goriely, Stephane},
  year = 2023,
  journal = {Eur. Phys. J. A},
  volume = {59},
  number = {6},
  pages = {131},
  doi = {10.1140/epja/s10050-023-01034-3},
  langid = {english}
}

@article{lemaire2005monte,
  title = {Monte {{Carlo}} Approach to Sequential Neutron Emission from Fission Fragments},
  author = {Lemaire, S. and Talou, P. and Kawano, T. and Chadwick, M. B. and Madland, D. G.},
  year = 2005,
  journal = {Phys. Rev. C},
  volume = {72},
  number = {2},
  pages = {024601},
  doi = {10.1103/PhysRevC.72.024601}
}

@article{litaize2015fission,
  title = {Fission Modelling with {{FIFRELIN}}},
  author = {Litaize, Olivier and Serot, Olivier and Berge, L{\'e}onie},
  year = 2015,
  journal = {Eur. Phys. J. A},
  volume = {51},
  number = {12},
  pages = {177},
  doi = {10.1140/epja/i2015-15177-9},
  langid = {english}
}

@article{marevic2026microscopic,
  title = {Microscopic Theory of Angular Momentum Distributions across the Full Range of Fission Fragments},
  author = {Marevi{\'c}, Petar and Schunck, Nicolas and Verriere, Marc},
  year = 2026,
  journal = {Phys. Rev. C},
  volume = {113},
  number = {1},
  pages = {014612},
  publisher = {American Physical Society},
  doi = {10.1103/yr2c-nvf3}
}

@book{ormand2021montea,
  title = {Monte {{Carlo Hauser-Feshbach}} Computer Code System to Model Nuclear Reactions: {{YAHFC}}},
  author = {Ormand, W. E.},
  year = 2021,
  publisher = {Lawrence Livermore National Security},
  address = {Livermore, Ca},
  collaborator = {{Lawrence Livermore National Laboratory}},
  langid = {english}
}

@article{pleasonton1973prompt,
  title = {Prompt {$\gamma$}-Rays Emitted in the Thermal-Neutron Induced Fission of {{233U}} and {{239Pu}}},
  author = {Pleasonton, Frances},
  year = 1973,
  journal = {Nucl. Phys. A},
  volume = {213},
  number = {2},
  pages = {413},
  doi = {10.1016/0375-9474(73)90161-9},
  langid = {english}
}

@article{schunck2016microscopic,
  title = {Microscopic Theory of Nuclear Fission: A Review},
  author = {Schunck, N. and Robledo, L. M.},
  year = 2016,
  journal = {Rep. Prog. Phys.},
  volume = {79},
  number = {11},
  pages = {116301},
  doi = {10.1088/0034-4885/79/11/116301}
}

@article{schunck2022theory,
  title = {Theory of Nuclear Fission},
  author = {Schunck, Nicolas and Regnier, David},
  year = 2022,
  journal = {Prog. Part. Nucl. Phys.},
  volume = {125},
  pages = {103963},
  doi = {10.1016/j.ppnp.2022.103963},
  langid = {english}
}

@article{talou2021fission,
  title = {Fission Fragment Decay Simulations with the {{CGMF}} Code},
  author = {Talou, P. and Stetcu, I. and Jaffke, P. and Rising, M. E. and Lovell, A. E. and Kawano, T.},
  year = 2021,
  journal = {Comput. Phys. Commun.},
  volume = {269},
  pages = {108087},
  doi = {10.1016/j.cpc.2021.108087},
  langid = {english}
}

@article{verbeke2018fission,
  title = {Fission {{Reaction Event Yield Algorithm FREYA}} 2.0.2},
  author = {Verbeke, J. M. and Randrup, J. and Vogt, R.},
  year = 2018,
  journal = {Comput. Phys. Commun.},
  volume = {222},
  pages = {263},
  doi = {10.1016/j.cpc.2017.09.006}
}

@article{wahl1988nuclearcharge,
  title = {Nuclear-Charge Distribution and Delayed-Neutron Yields for Thermal-Neutron-Induced Fission of $^{235}${U}, $^{233}${U}, and $^{239}${Pu} and for spontaneous fission of $^{252}${Cf}},
  author = {Wahl, Arthur C.},
  year = 1988,
  journal = {Atom. Data Nuc. Data Tab.},
  volume = {39},
  number = {1},
  pages = {1},
  doi = {10.1016/0092-640X(88)90016-2},
  langid = {english}
}

@preamble{ "\providecommand{\noopsort}[1]{} " }

\end{document}